\RequirePackage{lineno}
\documentclass[prd,twocolumn,amsmath,amssymb,showpacs,superscriptaddress,nofootinbib]{revtex4}
\usepackage{graphicx,epsfig}
\usepackage{dcolumn} % Align table columns on decimal point
\usepackage{bm,color}
\usepackage{multirow}
\usepackage[dvipdfm,CJKbookmarks=true,unicode,colorlinks,linkcolor=blue,anchorcolor=blue,citecolor=blue,pdfborder={0 0 0}]{hyperref}
\newcommand{\BR}{{\cal B}}
\newcommand{\diedecay}{\eta^\prime \rightarrow \pi^+\pi^- e^+e^-}
\newcommand{\dimudecay}{\eta^\prime \rightarrow \pi^+\pi^-\mu^+\mu^-}
\newcommand{\etap}{\eta^\prime}
\newcommand{\jpsi}{J/\psi}
\newcommand{\LL}{l^+l^-}
\newcommand{\EE}{e^+e^-}
\newcommand{\MM}{\mu^+\mu^-}
\newcommand{\pip}{\pi^+}
\newcommand{\pim}{\pi^-}

\newcommand{\g}{\gamma}
\newcommand{\ar}{\rightarrow}

\begin{document}
\linenumbers
\preprint{}

\title{\boldmath Measurement of $\etap\ar\pip\pim\EE$ and $\etap\ar\pip\pim\MM$} %$(l^{\pm}=e^{\pm}, \mu^{\pm})$}

\author{\small
M.~Ablikim$^{1}$, M.~N.~Achasov$^{6}$, O.~Albayrak$^{3}$,
D.~J.~Ambrose$^{39}$, F.~F.~An$^{1}$, Q.~An$^{40}$, J.~Z.~Bai$^{1}$,
R.~Baldini Ferroli$^{17A}$, Y.~Ban$^{26}$, J.~Becker$^{2}$
% ...
\\
\vspace{0.2cm}
(BESIII Collaboration)\\
\vspace{0.2cm}{\it
$^{1}$ Institute of High Energy Physics, Beijing 100049, People's Republic of China\\
$^{2}$ Bochum Ruhr-University, D-44780 Bochum, Germany\\
$^{3}$ Carnegie Mellon University, Pittsburgh, Pennsylvania 15213, USA\\
% ...
$^{6}$ G.I. Budker Institute of Nuclear Physics SB RAS (BINP), Novosibirsk 630090, Russia\\
% ...
$^{17}$ (A)INFN Laboratori Nazionali di Frascati, I-00044, Frascati, Italy; (B)INFN and University of Perugia, I-06100, Perugia, Italy\\
% ...
$^{26}$ Peking University, Beijing 100871, People's Republic of China\\
% ...
$^{39}$ University of Rochester, Rochester, New York 14627, USA\\
$^{40}$ University of Science and Technology of China, Hefei 230026, People's Republic of China\\
% ...
}\vspace{0.4cm}}

\begin{abstract}
  Based on a sample of 225.3 million $\jpsi$ events accumulated with
  the BESIII detector at the BEPCII, the decays of
  $\etap\ar\pip\pim\LL$ are studied via $\jpsi\ar\gamma\etap$. A clear
  $\etap$ signal is observed in the $\pip\pim\EE$ mass spectrum, and
  the branching fraction is measured to be
  $\BR(\etap\ar\pip\pim\EE)=(2.11\pm0.12~(stat.)\pm
  0.15~(syst.))\times10^{-3}$, which is in good agreement with
  theoretical predictions and the previous measurement, but is
  determined with much higher precision. No $\etap$ signal is found in
  the $\pip\pim\MM$ mass spectrum, and the upper limit is
  determined to be $\BR(\etap\ar\pip\pim\MM)<2.9\times10^{-5}$ at the
  90\% confidence level.
\end{abstract}


\maketitle

%% main text
\section{Introduction}\label{Introduction}

Since the $\etap$ was discovered in
1964~\cite{Kalbfleisch:1964ve,Goldberg:1964jn}, there has been
considerable interest in its decay both theoretically and
experimentally because of its special role in low energy scale
Quantum Chromodynamics (QCD) theory. Its main decay modes, including
hadronic and radiative decays, have been well
measured~\cite{PDG2012}, but the study of $\eta^\prime$ anomalous
decays is still an open field.

Recently, using the radiative decay $J/\psi\rightarrow \g\etap$ via
$\psi(3686)\rightarrow\pip\pim J/\psi$ as the source of $\eta^\prime$
mesons, CLEO~\cite{Naik:2009fj} reported the first observation of the
conversion decay $\etap\ar\pip\pim\EE$, which has been discussed for
many years based on
%VMD and CPT models {\color{red}}
the Vector Meson Dominance (VMD) model and Chiral Perturbation
Theory~\cite{Faessler:2000ef,Borasoy:2007pr,Petri:2010jy}.
Theoretically this decay is expected to proceed via a virtual photon
intermediate state, $\etap\ar\pip\pim \g^*\ar\pip\pim\EE$, and
provides a more stringent test of the theories since it involves
off-shell photons. In accordance with theoretical predictions, the
two prominent features expected for this decay are a peak with a
long tail just above $2m_{e}$ in the $e^+e^-$ ($M_{e^+e^-}$) mass
spectrum, and a dominant $\rho^0$ contribution in $M_{\pip\pim}$.
CLEO with limited statistics was unable to explore these
distributions, although their measured branching fraction,
$\BR(\etap\ar\pip\pim\EE)=(2.5^{+1.2}_{-0.9}\pm0.5)\times10^{-3}$~\cite{Naik:2009fj},
was consistent with predicted values around $2\times 10^{-3}$. In
addition, the search for $\etap\ar\pip\pim\MM$, which is predicted
to be lower by two order of magnitude, was also performed.  No
evident signal was observed, and the upper limit,
$\BR(\etap\ar\pip\pim\MM)<2.4\times10^{-4}$, at the 90\% confidence
level (C.L.), was determined.

At BESIII a sample of $(225.3\pm2.8)\times10^6$~\cite{Ablikim:2012vd}
$\jpsi$ events, corresponding to $1.2\times10^{6}$ $\etap$ events
produced through the radiative decay $\jpsi\ar\g\etap$, was collected
in 2009, and offers a unique opportunity to study $\etap$ decays. In
addition to $\etap\rightarrow \pip\pim l^+ l^-$,
$\etap\rightarrow\gamma\pip\pim$ is also studied in order to determine
the ratio of $\BR(\eta^\prime\rightarrow \pip\pim l^+ l^-)$ to
$\BR(\etap\ar\g\pip\pim)$. The advantage of measuring $\frac{
  \BR(\eta^\prime\rightarrow \pip\pim l^+
  l^-)}{\BR(\eta^\prime\rightarrow \gamma\pip\pim)}$ is that
uncertainties due to the number of $J/\psi$ events, tracking
efficiency from $\pi^{\pm}$ and the radiative photon detection
efficiency cancel.

\section{The experiment and monte carlo simulation} \label{Detector}

BEPCII is a double-ring $e^+e^-$ collider designed for a peak
luminosity of $10^{33}~\rm{cm}^{-2}\rm{s}^{-1}$ at the center of
mass energy of 3770 MeV. The cylindrical core of the BESIII detector
consists of a helium-gas-based drift chamber~(MDC) for charged track
and particle identification (PID) by $dE/dx$, a plastic scintillator
time-of-flight system~(TOF), and a 6240-crystal
CsI(Tl) Electromagnetic Calorimeter~(EMC) for electron
identification and photon detection. These components are all
enclosed in a superconducting solenoidal magnet providing a 1.0-T
magnetic field. The solenoid is supported by an octagonal
flux-return yoke with resistive-plate-counter muon detector modules
(MU) interleaved with steel. The geometrical acceptance for charged
tracks and photons is $93\%$ of $4\pi$, and the resolutions for
charged track momentum and photon energy at 1~GeV are $0.5\%$ and
$2.5\%$, respectively. More details on the features and capabilities
of BESIII are provided in Ref.~\cite{Ablikim:2009vd}.

The estimation of backgrounds and the determinations of detection
efficiencies are performed through Monte Carlo (MC) simulations. The
BESIII detector is modeled with the \textsc{geant}{\footnotesize
  4}~\cite{Agostinelli:2003hh,Allison:2006ve}.  The production of the
$J/\psi$ resonance is implemented with MC event generator
\textsc{kkmc}~\cite{Jadach:1999vf, Jadach:2000ir}, while the decays
are performed with \textsc{evtgen}~\cite{Ping2008}.  The possible
hadronic backgrounds are studied using a sample of $J/\psi$ inclusive
events in which the known decays of the $\jpsi$ are modeled with
branching fractions being set to the world average values in
PDG~\cite{PDG2012}, while the unknown decays are generated with the
\textsc{lundcharm} model~\cite{Chen:2000}.  For $\etap\ar\pip\pim\LL$
decays, a model~\cite{Zhangzy2012} based on theoretical calculations
using the vector meson dominant model with infinite-width corrections
and pseudoscalar meson mixing~\cite{Petri:2010jy} was developed.

\section{analysis} \label{Analysis}

\subsection{\bf $\etap\ar\pip\pim\LL$}

The final state in this analysis is $\g\pip\pim\LL$, with $l$ being an
electron or a muon. The charged tracks in the polar angle range
$|\cos\theta|<0.93$ are reconstructed from hits in the MDC. Good
charged tracks are required to pass within $\pm10~{\rm cm}$ of the
interaction point in the beam direction and $\pm1~{\rm cm}$ in the
plane perpendicular to the beam.  Photon candidates are reconstructed
by clustering the EMC crystal energies. The minimum energy is $25~{\rm
  MeV}$ for barrel showers ($|\cos\theta|<0.8$) and $50~{\rm MeV}$ for
end-cap showers ($0.86<|\cos\theta|<0.92$). To eliminate the showers
from charged particles, a photon must be separated by at least
$15^\circ$ from any good charged track. An EMC timing requirement is
used to suppress noise and energy deposits unrelated to the
event. Candidate events are required to contain exactly four good
charged tracks with zero net charge and at least one good photon. To
determine the species of the final state particles and select the best
photon when additional photons are found in an event, the combination
with the minimum value of $\chi^2_{\g\pip\pim\LL}$ is retained. Here
$\chi^{2}_{\g\pip\pim\LL}=\chi^{2}_{\rm 4C} +
\sum_{j=1}^{4}\chi^{2}_{\rm PID}(j)$ is the sum of the chi-square from
the four-constraint ($\rm 4C$) kinematic fit, and that from PID,
formed by combining TOF and $d\mathrm{E}/dx$ information of each
charged track for each particle hypothesis (pion, electron, or muon).
Events with $\chi^{2}_{\rm 4C}< 75$ are kept as $\g\pip\pim\LL$
candidates.  A $\rm 4C$ kinematic fit under the hypothesis of
$\g2(\pip\pim)$ is also performed, and $\chi^{2}_{\g2(\pip\pim)} >
\chi^{2}_{\g\pip\pim\LL}$ is required to reject possible
background events from $\jpsi\ar\g2(\pip\pim)$.

\begin{figure}[htbp]
    \centering
        \includegraphics[width=4.2cm]{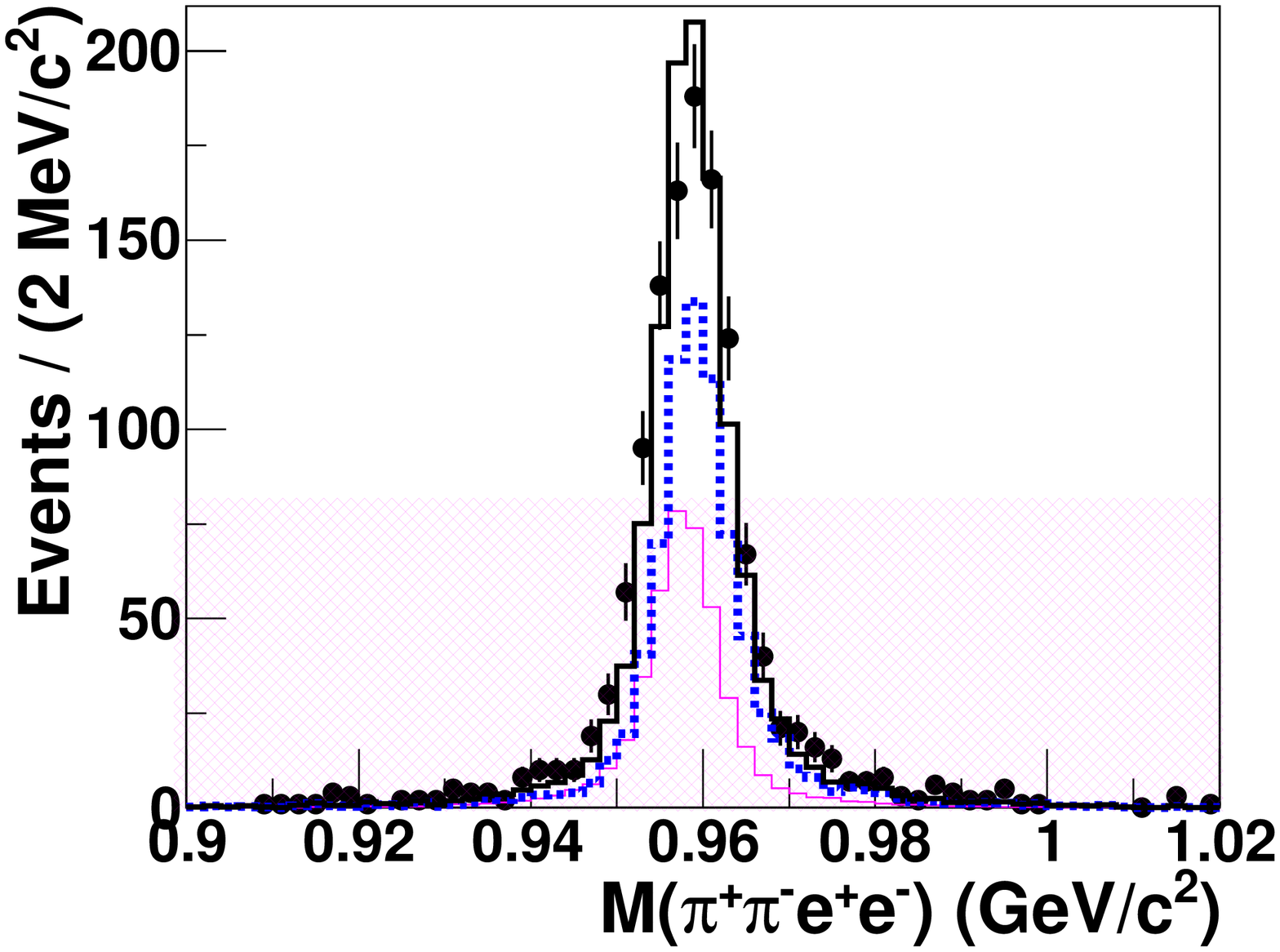}\put(-30,70){\bf \large~(a)} %_2pi2e_m2pi2e_da_sig_grho_arX_tmp.eps
        \includegraphics[width=4.2cm]{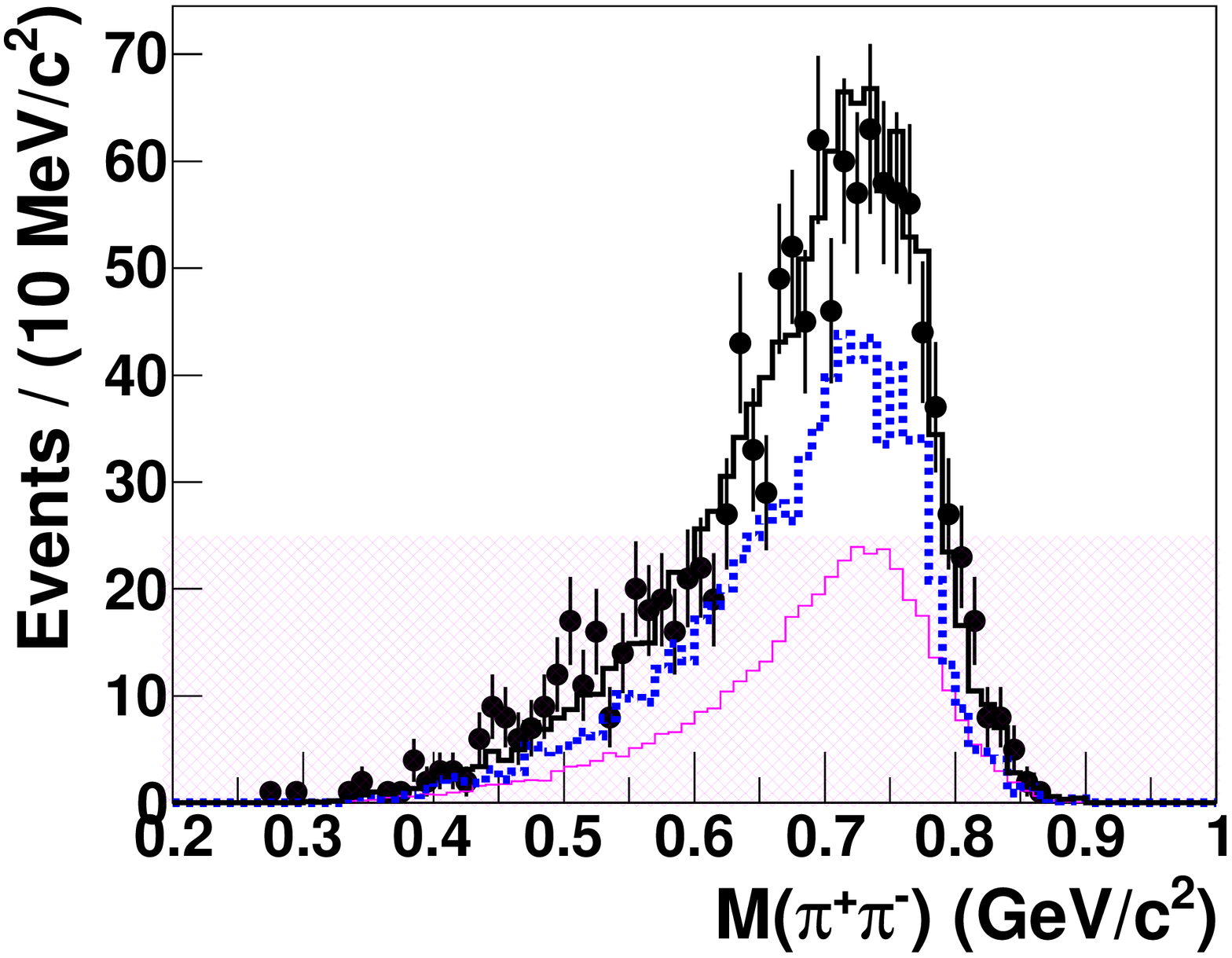}\put(-30,70){\bf \large~(b)}
        \caption{Kinematical distributions for the $\etap$ to
          $\pip\pim\EE$ decay: The invariant mass distributions of (a)
          ${\pip\pim\EE}$ and (b) $\pip\pim$. Dots with error bars
          represent the data; the shaded area is MC signal shape, the
          dashed histogram is the $\etap\ar\g\rho^{0}(\pip\pim)$ MC
          line shape, and the solid histogram is the sum of MC signal
          and MC background from $\etap\ar\g\rho^{0}(\pip\pim)$.  Both
          of these MC simulations are normalized to the yields found
          in Table I.}
    \label{fig:pipieekinematics}
\end{figure}

\begin{figure}[htbp]\centering
\includegraphics[width=0.45\textwidth]{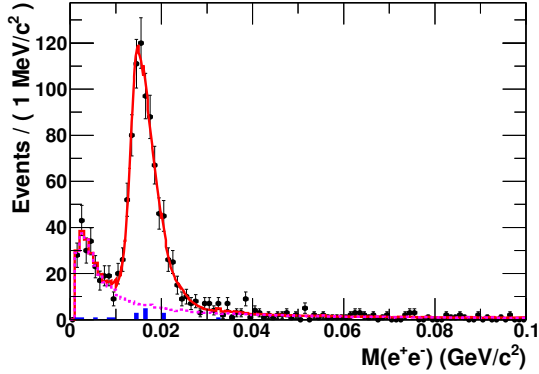}
\caption{The $\EE$ invariant mass spectrum of data (dots with error
  bars) after all selection criteria are applied. The solid line
  represents the fit result, the dotted histogram is the MC signal
  shape and the shaded histogram is background obtained from
  $\etap$ sideband events.}
    \label{fig:eekinematics}
\end{figure}

A very clear $\etap$ signal is observed in the $\pip\pim\EE$ invariant
mass distribution, shown in Fig.~\ref{fig:pipieekinematics}(a) after
the above event selection.  MC study shows that the dominant
background events come from $\jpsi\ar\g\etap$, $\etap\ar\g\pip\pim$
with the $\etap$ photon subsequently converted into an
electron-positron pair; this background is displayed as the dashed
histogram in Fig.~\ref{fig:pipieekinematics}(a). The di-pion invariant
mass distribution, which is shown in
Fig.~\ref{fig:pipieekinematics}(b), shows good agreement between data
and MC simulation.  Figure~\ref{fig:eekinematics} displays the ${\EE}$
mass spectrum after requiring $|M(\pip\pim\EE)-m(\etap)|<0.02$
GeV/$c^2$; the background from $\g\pip\pim$ conversions can be easily
distinguished.  The enhancement close to $e^+e^-$ mass threshold
corresponds to the signal from the $\etap\ar\pip\pim\EE$ decay, and
the clear peak around 0.015 GeV/$c^2$ comes from the background events
of $\etap\ar\g\pip\pim$ where the photon undergoes conversion to an
$e^+e^-$ pair and the electron (positron)'s momentum is improperly
reconstructed assuming that all the charged tracks are from the
interaction point. The background contributions of
$J/\psi\ar\pi^+\pi^-\pi^0$ and $J/\psi\ar\gamma\pi^+\pi^-\pi^0$ are
estimated from the $\eta^\prime$ sideband region (0.88 GeV/$c^2 <
M(\pip\pim\EE) <$ 0.90 GeV/$c^2$ or 1.02 GeV/$c^2 < M(\pip\pim\EE)<$
1.04 GeV/$c^2$).

To extract the $\etap\ar\pip\pim\EE$ events, a maximum likelihood
fit is performed on the observed $e^{+}e^{-}$ invariant mass
distribution with the signal shape described by the MC generator
specifically developed for this analysis, the dominant background
shape parameterized by a smooth function describing the $\g$
conversion events from $\etap\ar\g\pip\pim$, and the contribution
(17 events) obtained from $\etap$ sideband fixed in the fit to
account for the non-$\etap$ background. The fit, shown in
Fig.~\ref{fig:eekinematics}, yields $429\pm24$ $\pip\pim\EE$ events,
and the detection efficiency obtained from MC simulation is
$(16.94\pm0.08)$\%; both are summarized in
Table~\ref{tab:NumbersUsedInTheBranchingFractionCalculation}.

%\section{$\etap\ar\pip\pim\MM$}

\begin{figure}[htbp]\centering
\includegraphics[width=0.45\textwidth, height=5.5cm]{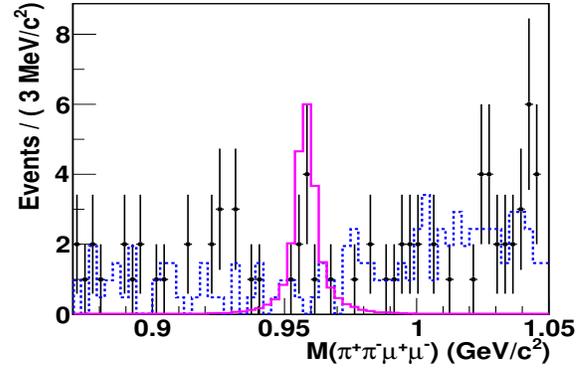}
\caption{The $\pip\pim\MM$ invariant mass distributions of data and MC
  simulation with all selection criteria applied. Dots with error bars
  represent the data, the solid histogram is MC signal, and the dashed
  line indicates inclusive MC.}
\label{fig:dimukinematics}
\end{figure}

Figure~\ref{fig:dimukinematics} shows the $\pip\pim\MM$ invariant mass
spectrum for candidates surviving all selection criteria. The
contribution from background events, mainly coming from
$\jpsi\ar\pi^0\pip\pim\pip\pim$ and $\jpsi\ar\g\pip\pim\pip\pim$ and
estimated with the inclusive MC $\jpsi$ events, is shown as the dashed
histogram. Although a few events accumulate in the $\etap$ mass
region, they are not significant.

%Assuming a flat prior for the $\etap$ signal, the Bayesian method is used
To determine the upper limit on the $\etap$ signal, a series of
unbinned maximum likelihood fits is performed to the mass spectrum of
${\pip\pim\MM}$ with an expected $\etap$ signal. In the fit, the line
shape of the $\etap$ signal is determined by MC simulation, and the
background is represented with a second-order Chebychev
polynomial. The likelihood distributions of the fit are taken as the
probability density function (PDF) directly.  The upper limit on the
number of signal events at the 90\% C.L. is defined as $N^{\rm U.L}$,
corresponding to the number of events at 90\% of the integral of the
PDF. The fit-related uncertainties on $N^{\rm U.L}$ are estimated by
using different fit ranges and different orders of the background
polynomial. The maximum one, $N^{\rm U.L}= 12$, and the detection
efficiency from MC simulation, $(35.47\pm 0.11)\%$, are used to
evaluate the upper limit on the branching fraction.

\subsection{\bf $\jpsi\ar\g\etap$, $\etap\ar\g\pip\pim$} \label{etap2grho}
%\boldmath
\begin{figure}[htpb]\centering
\includegraphics[width=0.45\textwidth]{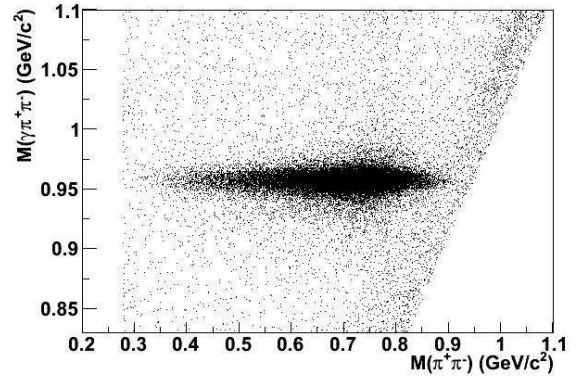}q
\caption{Scatter plot of $M({\g\pip\pim})$ versus $M({\pip\pim})$
for data.} \label{fig:m2pi_vs_mg2pipi_data}
\end{figure}

\begin{figure}[htpb]\centering
  \includegraphics[width=0.45\textwidth]{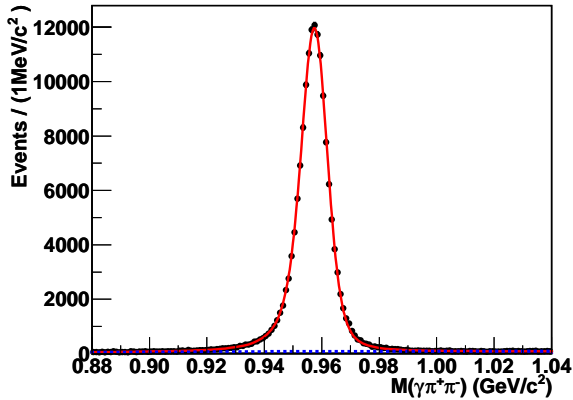}
  \caption{The $\g\pip\pim$ invariant mass spectrum for data after all
    selection criteria are applied. The solid curve is the fit
    result, and the dashed line represents the background polynomial.}
\label{fig:DIYMCshape_Conv_Gaussian_fit_arX}
\end{figure}

The final state is $\g\g\pip\pim$ for this mode. The charged track and
good photon selection are the same as those described above, but no PID is
applied in the event selection. A 4C kinematic fit is performed under
the hypothesis of $\jpsi\ar\pip\pim\g\g$, and $\chi_{\rm 4C}^{2} < 75$
is required.  For events with more than two photon candidates, the
combination with the minimum $\chi_{\rm 4C}^{2}$ is retained. To
reject background events with $\pi^0$ in the final state, the
invariant mass of the two photons is required to satisfy $M(\g\g)>$
0.16 GeV/$c^2$; this removes 94\% background while the efficiency loss
is only 0.73\%. The experimental signature of $\jpsi\ar\g\etap$
$(\etap\ar\g\pip\pim)$ is given by the radiative photon from $\jpsi$
decays, that carries a unique energy of 1.4 GeV. Consequently it is
easy to distinguish this photon from those from $\etap$ decays. In
this analysis, the combination of $\g\pip\pim$ invariant mass closest
to the $\etap$ mass is chosen to reconstruct the $\etap$.

\begin{figure}[htpb]\centering
\includegraphics[width=0.45\textwidth, height=5.8cm]{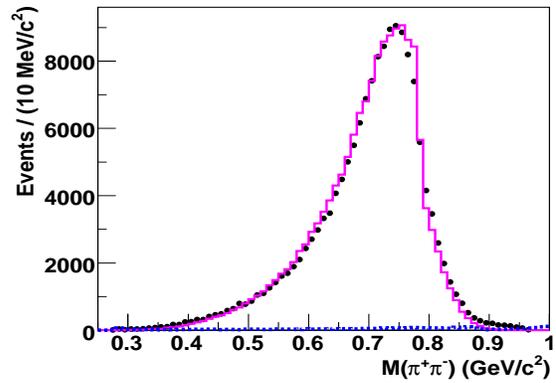} %grho_m2pi_data_MCs_VSS_DIY_arX.eps
\caption{The comparison of the simulated $\pip\pim$ mass spectrum with
  data. Dots with error bars are data within the $\etap$ region (
  [0.938, 0.978] GeV/$c^{2}$ ), the dashed histogram is background
  obtained from the $\etap$ sideband, and the solid histogram
  represents the MC simulation.
}\label{fig:grho_m2pi_data_MCs_VSS_DIY_arX}
\end{figure}

Figure~\ref{fig:m2pi_vs_mg2pipi_data} shows the scatter plot of
$M(\g\pip\pim)$ versus $M(\pip\pim)$ for the candidate events, where
the distinct $\etap-\rho^0$ band corresponds to the decay
$\etap\ar\g\pip\pim$. A very clean $\etap$ peak is observed in the
$M(\g\pip\pim)$ distribution, as displayed in
Fig.~\ref{fig:DIYMCshape_Conv_Gaussian_fit_arX}. The peak is fitted
with the MC simulated signal shape convolved with a Gaussian mass
resolution function to account for the difference in mass resolution
between data and MC simulations, plus a second-order Chebychev
polynomial background shape. The fit, shown as the smooth curve in
Fig.~\ref{fig:DIYMCshape_Conv_Gaussian_fit_arX} gives $158916\pm425$
$\eta^\prime\rightarrow \gamma \pip\pim$ events, and the detection
efficiency, $(45.39\pm0.07)\%$, is obtained from the MC simulation;
these are tabulated in
Table~\ref{tab:NumbersUsedInTheBranchingFractionCalculation}. In the
simulation of $\eta^\prime\rightarrow \gamma \pip\pim$, since the
resonant contribution from $\rho^0\rightarrow \pi^+\pi^-$ is
insufficient to describe the data, the non-resonant contribution
(known as the "box anomaly") is also included using a decay rate
formula~\cite{dacay_rate} deduced from the ones used in
Refs.~\cite{theory1, theory2, theory3}. With the parameters tuned
with data, the comparison of the simulated dipion mass spectrum to
data in Fig.~\ref{fig:grho_m2pi_data_MCs_VSS_DIY_arX} shows good
agreement.

\begin{table}
    \centering
    \caption{Numbers used in the branching fraction calculations: the fitted signal yields, $N$ (or 90\% C.L. upper limit); the detection efficiency, $\epsilon$.}
        \begin{tabular}{lccc}
        \hline
        $\etap$ decay mode        &$\epsilon$ (\%)             &$N$    \\
        \hline
        $\pip\pim\EE$                 &16.94~$\pm$ 0.08   &429~$\pm$ 24 \\
        $\pip\pim\MM$               &35.47~$\pm$ 0.11   &$<$ 12      \\
        $\g\rho^0 (\pip\pim)$    &45.39~$\pm$ 0.07    &158916~$\pm$ 425  \\%\multicolumn{2}{c}{\hspace{5mm} \hspace{3mm}}
        \hline
        \\
        \end{tabular}
    \label{tab:NumbersUsedInTheBranchingFractionCalculation}
\end{table}

\section{Systematic errors}

In the measurement of the ratio of the branching fractions, the
possible systematic error sources and the corresponding
contributions are discussed in detail below.
\begin{itemize}
\item Form factor uncertainty. In the MC generator used to determine
  the detection efficiency of $\etap\ar\pip\pim\LL$, the VMD factor defined for the hidden gauge model is
  introduced to account for the contribution from the $\rho^0$
  meson. The detection efficiency dependence is evaluated by replacing
  the factor above with the modified VMD factors denoted in
  Ref.~\cite{Petri:2010jy}. The maximum change of the detection
  efficiencies is assigned as the systematic error, which is listed in
  Table~\ref{tab:sys}.

\item MDC tracking efficiency. Since the systematic errors for the two
  charged pions cancel by measuring the relative branching fraction of
  $\etap\ar\pip\pim\LL$ and $\etap\ar\g\pip\pim$, only the systematic
  error caused by the MDC tracking from the leptonic pairs need be
  considered. As the momenta of the two charged leptons are quite low,
  it is difficult to select a pure sample from data. In this analysis
  the MDC tracking uncertainty of charged pions at low momentum is
  determined and used to estimate that of the leptons by
  reweighting
%with the re-weighting method.
  in accordance with their momenta. The data sample of
  $\jpsi\ar\g\etap$, $\etap\ar\g\pip\pim$ is used to evaluate the
  data-MC difference of pions at low momentum and finally the MDC
  tracking uncertainty is estimated to be 2.1\% for electrons and
  1.6\% for muons, where the dominant contribution is from the
  momentum region below 200 MeV/$c$. Therefore 4.2\% and 3.2\% are
  taken as the systematic errors on the tracking efficiency for the
  channels with $e^+e^-$ and $\mu^+\mu^-$, respectively, in the final
  states.

\item Photon detection efficiency. The photon detection efficiency is
  studied with three independent decay modes, $\psi(2S)\ar\pi^+\pi^-
  J/\psi$ $(J/\psi\ar\rho^0\pi^0)$, $\psi(2S)\ar\pi^+\pi^- J/\psi$
  $(J/\psi\ar l^+ l^-)$ and
  $J/\psi\ar\rho^0\pi^0$~\cite{photondec}. The results indicate that
  the difference between the detection efficiency of data and MC
  simulation is within 1\% for each photon. Since the uncertainty from
  the radiative photons cancel by measuring the relative branching
  fraction of $\etap\ar\pip\pim\LL$ and $\etap\ar\g\pip\pim$, 1\% is
  taken to be the systematic error from the photon in $\etap$ decaying
  into $\g\pip\pim$.

\item Particle ID. The study of the particle ID efficiency of the pion
  is performed using the clean control sample of
  $\jpsi\ar\pip\pim\pi^0$, and indicates that the pion particle ID
  efficiency for data agrees within 1\% of that of the MC simulation
  in the pion momentum region. The particle ID efficiency of the
  electron was checked with radiative Bhabha events, and the difference
  between data and MC simulation is found to be 1\%. In this analysis,
  4\% is taken as the systematic error from the particle ID efficiency
  of the four charged tracks in $\etap$ decaying into $\pip\pim\LL$.

\item Kinematic fit. The clean sample $\jpsi\ar\phi\eta$ ($\phi\ar K^+
  K^-$, $\eta\ar\pip\pim\pi^{0}$) selected without a kinematic fit is
  used to estimate the systematic error associated with the 4C
  kinematic fit. The difference between data and MC is determined to
  be $(0.47\pm 1.45)$\%, with $\chi^2<75$. In this paper, 1.9\% is
  taken to be the systematic error from the kinematic fit for the
  analyzed decays of $\jpsi\ar\g\etap$ $(\etap\ar\pip\pim\LL)$. For
  $\jpsi\ar\g\etap$, $\etap\ar\g\pip\pim$ channel, the 4C kinematic
  fit uncertainty is estimated to be less than 0.7\% using the control
  sample $\jpsi\ar\rho\pi$. Thus, the error from kinematic fit is,
  2.0\%, the sum of them added in quadrature.
%{\color{red} }
\item Background uncertainty. Studies have shown that the mass
  resolution of $\g\pip\pim$, as simulated by the MC, is
  underestimated. To evaluate the systematic effect associated with
  this, the invariant mass of $\g\pip\pim$ in the MC sample is smeared
  with a Gaussian function, where the width of this Gaussian is
  floated in the fit.  The change of the result, 0.9\%, is assigned to
  be the systematic error.

\item $\etap$ mass window requirement. Another source of systematic
  uncertainty is the requirement on the $\etap$ mass window selection
  $|M({\pi^+\pi^- e^+e^-})-m({\etap})|<0.02$ GeV/$c^2$. The
  uncertainty is studied using a looser requirement of 0.90 GeV/$c^2$
  $< M(\pi^+\pi^- e^+e^-)<1.02$ GeV/$c^2$, and an uncertainty of 2.0\%
  is assigned for this item.

\item Uncertainty of the number of $\etap\ar\g\pip\pim$ events
  ($N_{\etap\ar\g\pip\pim}$). The uncertainty from this item, 0.5\%,
  contains the error due to the $\pi^0$ veto cut ($M(\g\g)>$ 0.16
  GeV/$c^2$) and the fit-related error.
\end{itemize}

Except for the systematic uncertainties studied above, a small
uncertainty due to the statistical error of the efficiencies in
$\etap\ar\pip\pim\LL$ and $\etap\ar\g\pip\pim$ is also considered; all
errors are summarized in Table~\ref{tab:sys}.  The total systematic
error is the sum of them added in quadrature.

\begin{table}
    \centering
    \caption{Impact (in \%) of the systematic uncertainties on the measured branching fractions.}
\begin{tabular}{lcc}
\hline
Sources &$\diedecay$ &$\dimudecay$\\
\hline
Form factor uncertainty &0.2   &0.3 \\
MDC tracking            &4.2   &3.2 \\
Photon detection        &1.0   &1.0 \\
PID                     &4.0   &4.0 \\
4C kinematic fit        &2.0   &2.0 \\
Background uncertainty  &0.9   &--  \\
$\etap$ mass window     &2.0   &-- \\
$N_{\etap\ar\g\pip\pim}$  &0.5   &0.5\\
MC statistics             &0.6   &0.4 \\
\hline
Total &6.6  &5.6  \\
\hline
\end{tabular}
    \label{tab:sys}
\end{table}

\section{Results}\label{Results}

The ratio (upper limit) of $\BR(\etap\ar\pip\pim\LL)$ to
$\BR(\etap\ar\g\pip\pim)$ is calculated with
\begin{equation*}
\frac{\BR(\etap\ar\pip\pim\LL)}{\BR(\etap\ar\g\pip\pim)}=
\frac{N_{\etap\ar\pip\pim\LL}/\epsilon_{\etap\ar\pip\pim\LL}}{N_{\etap\ar\g\pip\pim}/\epsilon_{\etap\ar\g\pip\pim}},
\end{equation*}
where $N_{\etap\ar\pip\pim\LL}$ and $N_{\etap\ar\g\pip\pim}$ are the
observed events (or the 90\% C.L. upper limit) of
$\etap\ar\pip\pim\LL$ and $\etap\ar\g\pip\pim$, and
$\epsilon_{\etap\ar\pip\pim\LL}$ and $\epsilon_{\etap\ar\g\pip\pim}$
are the corresponding detection efficiencies. With the numbers given
in Table~\ref{tab:NumbersUsedInTheBranchingFractionCalculation}, the
ratio $\frac{\BR(\etap\ar\pip\pim\EE)}{\BR(\etap\ar\g\pip\pim)}$ is
determined to be $(7.2\pm0.4~(stat.)\pm 0.5 ~(syst.))\times10^{-3}$,
where the first error is the statistical error from
$N_{\etap\ar\pip\pim\LL}$ and $N_{\etap\ar\g\pip\pim}$.  To
calculate the upper limit, the systematic error is taken into
account by a factor of $\frac{1}{1-\delta_{syst}}$. Therefore the
upper limit, $1.0\times10^{-4}$, on the ratio
$\frac{\BR(\etap\ar\pip\pim\EE)}{\BR(\etap\ar\g\pip\pim)}$ is given
at the 90\% confidence level.

\section{Summary}\label{Summary}
The measurements of $\etap\ar\pip\pim\LL$, $l^{\pm} =
(e^{\pm},\mu^{\pm})$ are performed using the sample of 225.3 million
$\jpsi$ events collected with the BESIII detector. A clear signal is
observed in the invariant mass spectrum of $\pip\pim\EE$, and the
ratio $\frac{\BR(\etap\ar\pip\pim\EE)}{\BR(\etap\ar\g\pip\pim)}$ is
determined to be $(7.2\pm0.4~(stat.)\pm 0.5~(syst.))\times10^{-3}$.
Using the PDG world average of $\BR(\etap\ar\g\pip\pim)$ and its
uncertainty~\cite{PDG2012}, the branching fraction is measured to be
$\BR(\etap\ar\pip\pim\EE) = (2.11\pm 0.12~(stat.)\pm
0.15~(syst.))\times10^{-3}$ which is consistent with the theoretical
predictions and previous measurement, but with the precision improved
significantly. The mass spectra of $\pip\pim$ and $e^+e^-$ are also
consistent with the theoretical predictions that $M_{\pip\pim}$ is
dominated by $\rho^0$, and $M_{e^{+}e^{-}}$ has a peak just above
2$m_{e}$ with a long tail.  No evidence for $\etap$ decaying into
$\pip\pim\MM$ is found, and an upper limit of $1.0\times10^{-4}$ on
the ratio of
$\frac{\BR(\etap\ar\pip\pim\MM)}{\BR(\etap\ar\g\pip\pim)}$ is obtained
at the 90\% confidence level. The corresponding branching fraction
upper limit of $\etap\ar\pip\pim\MM$ is
$\BR(\etap\ar\pip\pim\MM)<2.9\times10^{-5}$.

\section{Acknowledgment}
\label{sec:Acknowledgement}

The BESIII collaboration thanks the staff of BEPCII and the
computing center for their hard efforts. This work is supported in
part by the Ministry of Science and Technology of China under
Contract No. 2009CB825200; National Natural Science Foundation of
China (NSFC) under Contracts Nos. 10625524, 10821063, 10825524,
10835001, 10935007, 10979033, 10979012, 11175189, 11125525,
11235011; Joint Funds of the National Natural Science Foundation of
China under Contracts Nos. 11079008, 11179007; the Chinese Academy
of Sciences (CAS) Large-Scale Scientific Facility Program; CAS under
Contracts Nos. KJCX2-YW-N29, KJCX2-YW-N45; 100 Talents Program of
CAS; German Research Foundation DFG under Contract No. Collaborative
Research Center CRC-1044; Istituto Nazionale di Fisica Nucleare,
Italy; Ministry of Development of Turkey under Contract No.
DPT2006K-120470; U. S. Department of Energy under Contracts Nos.
DE-FG02-04ER41291, DE-FG02-05ER41374, DE-FG02-94ER40823; U.S.
National Science Foundation; University of Groningen (RuG) and the
Helmholtzzentrum fuer Schwerionenforschung GmbH (GSI), Darmstadt;
WCU Program of National Research Foundation of Korea under Contract
No. R32-2008-000-10155-0. This paper is also supported by the
Natural Science Foundation of Shandong Province, China under
Contracts Nos. 2009ZRB02465.

%This paper is also supported by the NSFC under Contract Nos. 10979033, 10979012, 11175189; Natural Science Foundation of Shandong
%Province, China under Contracts Nos. 2009ZRB02465.

%\bibliographystyle{model1-num-names} %%Nmbered references with article and chapter titles, listed in order of citation
%\bibliography{}
%\bibliography{literature}

\begin{thebibliography}{18}

%\cite{Jaffe:1976ig} %%%% reference 1
\bibitem{Kalbfleisch:1964ve}
  G.~R.~Kalbfleisch {\it et al.},
  Phys.\ Rev.\ Lett.\  {\bf 12}, 527 (1964).

\bibitem{Goldberg:1964jn}
  M.~Goldberg {\it et al.},
  Phys.\ Rev.\ Lett.\  {\bf 12}, 546 (1964).

\bibitem{PDG2012}
  J.~Beringer {\it et al.},
  Phys.\ Rev.\  D {\bf 86}, 010001 (2012).

\bibitem{Naik:2009fj}
  P.~Naik {\it et al.},
  Phys.\ Rev.\ Lett.\  {\bf 102}, 061801 (2009).

\bibitem{Faessler:2000ef}
  A.~Faessler, C.~Fuchs, M.~I.~Krivoruchenko,
  Phys.\ Rev.\ C {\bf 61}, 035206 (2000).

\bibitem{Borasoy:2007pr}
  B.~Borasoy, R.~Nissler,
  Eur.\ Phys.\ J.\ A {\bf 33}, 95 (2007).

\bibitem{Petri:2010jy}
  T.~Petri,
  arXiv:1010.2378 [nucl-th].

\bibitem{Ablikim:2012vd}
  M.~Ablikim {\it et al.},
  Chin.\ Phys.\ C {\bf 36}, 915 (2012).

\bibitem{Ablikim:2009vd}
  M.~Ablikim {\it et al.},
  Nucl.\ Instrum.\ Meth.\ A {\bf 614}, 345 (2010).

\bibitem{Agostinelli:2003hh}
  S.~Agostinelli {\it et al.},
  Nucl.\ Instrum.\ Meth.\ A {\bf 506}, 250 (2003).

\bibitem{Allison:2006ve}
  J.~Allison, K.~Amako, J.~Apostolakis, H.~Araujo, P.~Dubois {\it et
  al.},
  IEEE Trans.\ Nucl.\ Sci.\ {\bf 53}, 270 (2006).

\bibitem{Jadach:1999vf}
  S.~Jadach, B.~Ward, Z.~Was,
  Comput.\ Phys.\ Commun.\ {\bf 130}, 260 (2000).

\bibitem{Jadach:2000ir}
  S.~Jadach, B.~Ward, Z.~Was,
  Phys.\ Rev.\  D {\bf 63}, 113009 (2001).

\bibitem{Ping2008}
  R. G.~Ping {\it et al.},
  Chin.\ Phys.\ C {\bf 32}, 599 (2008).

\bibitem{Chen:2000}
  J. C. Chen {\it et al.},
  Phys.\ Rev.\  D {\bf 62}, 1 (2000).

\bibitem{Zhangzy2012}
  Z.~Y.~Zhang, L.~Q.~Qin, S.~S.~Fang,
  Chin.\ Phys.\ C {\bf 36}, 926 (2012).

\bibitem{dacay_rate}
$\frac{d\Gamma}{dm}\propto k^3_{\gamma}q^3_{\pi}(m)|BW_{\rho}^{GS}(1+\delta\frac{m^2}{m_{\rho}^2}BW_{\omega})+\beta|^2$,  where $k_{\gamma}$ is the photon energy and $q_{\pi}(m)$ is the momentum of pion in the $\pip\pim$ rest frame.
$BW_{\rho}^{GS}$ is the Breit-Wigner dsitribution in GS parameterazation~\cite{theory4}.
%which can describe the $\rho$ resonance well in $e^+e^-\ar\pip\pim$ experiments~\cite{theory2}.
$|\delta|$ represents the contribution from $\omega$ resonance and the complex phase of $\delta$ represents the interference between $\omega$ and $\rho(770)$ resonance. $m_{\rho}$ is the mass of the $\rho(770)$ resonance. $\beta$ represents the contribution from the non-resonance.
%$\frac{E_{\etap}}{F_{\etap}}$ is constant ratio, which represents the contribution from box anomaly.

%\bibitem{formula}  please note down the formula here
\bibitem{theory1}  A. Abele et al., Phys. Lett., B{\bf 402}, 195 (1997). %Crystal Barrel Collab.,
\bibitem{theory2}  R. R. Akhmetshin et al., Phys. Lett., B{\bf 527}, 161 (2002).
\bibitem{theory3}  M. Benayoun et al., Z. Phys. C{\bf 58}, 31 (1993).
%\bibitem{Chen2007} H.~X.~Chen, Int.\ J.\ Mod.\ Phys.\ A {\bf 22}, 637 (2007).

\bibitem{photondec}
  M.~Ablikim {\it et al.},
  Phys.\ Rev.\  D {\bf 83}, 112005 (2011).

\bibitem{theory4} G.~J.~Gounaris, J.~J.~Sakurai, Phys.\ Rev.\ Lett.\  {\bf 21}, 244 (1968).

\end{thebibliography}

\end{document}